\documentclass[journal]{IEEEtran}
\usepackage{amssymb}
\DeclareUnicodeCharacter{2061}{}
\usepackage{epstopdf}   
\usepackage{epsfig}
\usepackage{graphicx}
\usepackage{comment}
\usepackage{enumerate}
\usepackage[utf8]{inputenc}
\usepackage{lettrine}
\usepackage{hyperref}
\usepackage{algorithm}
\usepackage{tikz}
\usepackage{algpseudocode}
\usepackage{epsfig}
\usepackage{bibentry}
\usepackage{amsmath}
\usepackage{url}
\usepackage{rotating}
\usepackage{array}
\usepackage{stfloats}
\usepackage{blindtext}
\usepackage{fancyhdr}
\usepackage{graphicx}
\usepackage{caption}
\usepackage{subcaption}
\usepackage{verbatim}
\usepackage{xcolor}
\usepackage{url}
\usepackage{comment}
\usepackage{lettrine}
\usepackage{cite}
\usepackage[normalem]{ulem}
\usepackage{amsmath}
\usepackage{pifont}
\newcommand{\cmark}{\text{\ding{51}}}
\newcommand{\xmark}{\text{\ding{55}}}
\useunder{\uline}{\ul}{}
\makeatletter
\g@addto@macro{\UrlBreaks}{\UrlOrds}
\makeatother
\ifCLASSINFOpdf 
\begin{document}

\title{Generative Federated Learning for Smart Prediction and Recommendation Applications\\
\author{Anwesha Mukherjee, Rajkumar Buyya, \IEEEmembership{Fellow, IEEE}
\thanks{Anwesha Mukherjee is with the Quantum Cloud Computing and Distributed Systems Laboratory, School of Computing and Information Systems, The University of Melbourne, Melbourne, Australia. She is also with the Department of Computer Science, Mahishadal Raj College, Mahishadal, West Bengal, India (e-mail: anweshamukherjee@mail.mrc.ac.in, anweshamukherjee2011@gmail.com).}
\thanks{Rajkumar Buyya with the Quantum Cloud Computing and Distributed Systems Laboratory, School of Computing and Information Systems, The University of Melbourne, Melbourne, Australia (e-mail: rbuyya@unimelb.edu.au)}}
}

\maketitle

\begin{abstract}
This paper proposes a generative adversarial network and federated learning-based model to address various challenges of the smart prediction and recommendation applications, such as high response time, compromised data privacy, and data scarcity. The integration of the generative adversarial network and federated learning is referred to as Generative Federated Learning (GFL). As a case study of the proposed model, a heart health monitoring application is considered. The realistic synthetic datasets are generated using the generated adversarial network-based proposed algorithm for improving data diversity, data quality, and data augmentation, and remove the data scarcity and class imbalance issues. In this paper, we implement the centralized and decentralized federated learning approaches in an edge computing paradigm. In centralized federated learning, the edge nodes communicate with the central server to build the global and personalized local models in a collaborative manner. In the decentralized federated learning approach, the edge nodes communicate among themselves to exchange model updates for collaborative training. The comparative study shows that the proposed framework outperforms the existing heart health monitoring applications. The results show that using the proposed framework (i) the prediction accuracy is improved by $\sim$12\% than the conventional framework, and (ii) the response time is reduced by $\sim$73\% than the conventional cloud-only system.
\end{abstract}

\begin{IEEEkeywords}
Heart health monitoring, federated learning, response time, energy consumption.
\end{IEEEkeywords}

\section{Introduction}
\textcolor{black}{Machine learning (ML) and the Internet of Things (IoT) have become two principal technologies for smart applications, such as healthcare, agriculture, and transportation systems. For large-scale data analysis, deep learning (DL) plays an important role. However, fast and accurate analysis of the collected data is important for early detection, especially for hard deadline applications such as healthcare. Early diagnosis is a vital parameter for healthcare to prevent fatal incidents. Accurate health status prediction at lower response time is highly significant in Internet of Health Things (IoHT), which refers to the application of Internet of Things (IoT) for health monitoring \cite{mukherjee2020fogioht}. Heart health monitoring \cite{ibrahim2024end, butkow2023heart} is one of the significant issues in smart healthcare, where continuous monitoring of blood pressure, blood sugar, maximum heart rate, chest pain, ECG, etc., is required. Further, privacy is another concern for personal health data, which is sensitive and confidential. In conventional IoT-based systems, health data analysis and storage take place inside the cloud servers, which causes a major concern regarding data privacy. Federated learning (FL) has come as an emerging technology to provide privacy-aware data analysis through collaborative training and local data analysis \cite{nguyen2021federated, bera2024flag}. In an FL-based edge computing environment, the edge nodes perform local data analysis, and no data is shared with the cloud \cite{bera2024flag}. The participating nodes either communicate with the server to undergo a collaborative learning process (centralized FL) or communicate among themselves to perform a collaborative learning process (decentralized FL) to develop a global model \cite{bera2024flag, bera2024fedchain}.} 
\par
\textcolor{black}{Though FL prevents the sharing of raw data and protects data privacy, several issues still remain, such as limited data samples, class imbalance in the dataset, which affects the prediction accuracy of the generated model. As the data are non-independent and non-identically distributed (non-IID) across the edge nodes, building an accurate prediction model through aggregation becomes difficult. Hence, a model is required that will address these issues and build an accurate prediction model with a low response time to prevent fatal incidents. Generative artificial intelligence approaches can learn the patterns of the input data and generate data with similar characteristics. GAN \cite{goodfellow2020generative, jabbar2021survey, saxena2021generative} models identify the patterns of the given input data and accordingly generate synthetic data of similar characteristics. This paper proposes a generative adversarial network (GAN) and FL-based edge-cloud system with a case study of heart health monitoring application.} 

\subsection{Motivation and Contributions}
\textcolor{black}{Accurate and privacy-aware data analysis at a low response time is one of the major challenges of smart systems. Though, FL protects data privacy, the non-IID dataset handling is a major challenge. The data scarcity and class imbalance issues hampers the prediction accuracy of the model. The motivation of this work is to address these issues and provide an accurate prediction model with minimal response time. The key contributions of this paper are summarized as follows:}
\begin{itemize}
    \item A GAN- and FL-based model is proposed for smart systems with a case study of a smart health monitoring application. The integration of the generative adversarial network and federated learning is referred to as Generative Federated Learning (GFL).
    \item GAN is used to generate realistic data to resolve the data scarcity and class imbalance issues, especially for the non-IID datasets across the clients.
    \item An FL-based model is developed, where the edge nodes communicate with the cloud or among themselves to perform a collaborative training process to build a global model. Both the centralized federated learning (CFL) and decentralized federated learning (DFL) frameworks are experimentally implemented. The Gated Recurrent Unit (GRU) is used as the underlying deep learning (DL) model to capture the sequential data pattern.
    \item The prediction accuracy, training time, response time, and energy consumption are measured and compared with the conventional systems to showcase the efficacy of the proposed approach.
\end{itemize}

The rest of the paper is organized as follows: The existing approaches are discussed in Section \ref{rel}. Section \ref{pro} presents the proposed system model along with the algorithms. The performance of the proposed framework is analyzed in Section \ref{per}. Finally, we conclude in Section \ref{con}.

\section{Related Work}
\label{rel}
\textcolor{black}{IoT, ML, and DL play important roles in smart applications. For privacy-aware data analysis without raw data transmission, FL has come \cite{bera2024flag, bera2024fedchain}. Smart healthcare is one of the significant domain where various research works were carried out. As health-related data is confidential and highly sensitive, FL-based healthcare has become an emerging area of research. Table \ref{tab:1} presents a feature-based comparative study between the proposed and existing FL-based healthcare systems.}
\par
\textit{ML, DL, and FL in healthcare:}
The use of ML and DL in smart healthcare was explored in various works \cite{yigit2024machine, lv2022deep, li2021comprehensive, guo2022federated}. The use of ML, IoT, and edge computing in healthcare was explored in \cite{alnaim2023machine}. In \cite{nguyen2022federated, li2021federated}, the use of FL in smart healthcare was discussed. In \cite{yaqoob2023hybrid}, the use of FL in cardiovascular disease prediction was explored. The authors used modified artificial bee colony optimization with support vector machine (MABC-SVM) in their scheme \cite{yaqoob2023hybrid}. In \cite{ghazal2025generative}, FL was used with large and small language models for healthcare applications. For medical data processing in smart healthcare, FL was used in \cite{guo2022federated}. Though the use of CFL was explored in most of the existing works, the use of DFL was not explored. In CFL, the overhead on the cloud is high as the aggregation takes place inside the cloud after receiving all updates from the clients. In such a case, DFL can be used, where multiple devices collaborate among themselves to develop a global model. 
\par
\textit{GANs in healthcare:}
For healthcare data analysis, GANs have become popular for dealing with data scarcity issues \cite{vaccari2021generative, indhumathi2022healthcare}. GANs allow for generating synthetic data that resembles the original data \cite{goodfellow2020generative}. GAN is composed of a generator and a discriminator. The generator is used to produce synthetic data and the discriminator is used to verify the genuineness of the generated data. For alzheimer's disease classification, GAN was used in \cite{tufail2024deep}. In \cite{purandhar2022classification}, the authors used GAN for healthcare data classification. In \cite{la2022deep}, hyperspectral lesion images were generated using GAN for skin cancer diagnosis. For lung cancer data analysis, GAN was used to generate synthetic data \cite{gonzalez2021generative}. A GAN-based method was proposed in \cite{indhumathi2022healthcare} for healthcare data generation. Though the existing approaches on GAN dealt with the data scarcity issue, the use of CFL and DFL with GAN was not explored to achieve privacy-awareness, low response time, and high accuracy simultaneously.
\par
\textit{Edge and fog computing in healthcare:}
Response time is one of the important parameter for real-time applications. For hard-deadline applications like healthcare, low response time is highly desirable. The conventional cloud-based systems suffer from high response time, high bandwidth requirement, interruption in connectivity, etc. Edge computing has come as a solution to address these challenges by bringing resources at network edge. Fog computing also reduces latency by allowing the intermediate nodes to participate in data processing. An edge computing-based healthcare system was discussed in \cite{singh2023edge}. In \cite{mukherjee2020fogioht}, fog computing was used for IoHT applications. In \cite{amin2020edge}, the use of IoT, edge and fog computing in healthcare was discussed. The benefits and challenges of edge-based smart healthcare were discussed in \cite{abdellatif2019edge}. For 5G-based smart healthcare systems, wireless body area network with edge computing was discussed in \cite{bishoyi2021enabling}. However, edge/fog computing alone cannot handle the issue of data scarcity as well as privacy protection. 
\par
From the above discussion, we observe that an integration of GAN, FL, and edge computing is required to achieve:
\begin{itemize}
    \item Low response time
    \item Accurate prediction
    \item Data privacy protection
    \item Sufficient data samples for proper training to resolve data scarcity and class imbalance issues, especially for non-IID datasets
\end{itemize}
To fulfill all of these, we propose a GAN and FL-based edge computing framework for smart healthcare.
\par
\textit{Comparison with existing approaches:} The comparison of the proposed work with existing FL-based healthcare systems are presented in Table \ref{tab:1}. As we observe most of the existing approaches used CFL, whereas we implement CFL as well as DFL architectures. We observe from Table \ref{tab:1} that the proposed approach uses FL, GAN, and edge computing. GAN is used to deal with data scarcity and improve data diversity. Further, as the generated synthetic dataset is used for analysis, the concern regarding disclosure of personal data is also resolved. For privacy-aware data analysis, FL is used in the proposed work. FL with edge computing is used to reduce the response time in healthcare service provisioning. As the response time is vital for hard-deadline applications like healthcare, and energy consumption is significant for a sustainable system, we not only determine the prediction accuracy, precision, recall, and F-score, but also the time and energy consumption while evaluating the performance of the proposed framework. Hence, in comparison with the existing schemes, this is observed that the proposed approach is unique and better. 
\begin{table} 
\caption{Comparison among the proposed and existing FL-based healthcare systems}
    \centering
    \begin{tabular}{|c|c|c|c|c|c|}
        \hline
         Work   &  FL- &  GAN &  Edge-  &  Time/ & Energy \\
          &  based	& is  &  based & latency is & is \\
          & system &  used &system& measured & measured\\
          \hline
          Guo & \cmark & \xmark & \xmark & \xmark & \xmark\\
         et al. \cite{guo2022federated}  & & & & &\\
        \hline
         Li & \cmark & \xmark & \xmark & \cmark & \xmark\\
         et al. \cite{li2021federated}  & & & & &\\
        \hline
         Yaqoob  & \cmark & \xmark & \xmark & \xmark & \xmark\\
         et al. \cite{yaqoob2023hybrid}  & & & & &\\
         \hline
          Wu & \cmark & \cmark & \cmark & \cmark & \xmark\\
         et al. \cite{wu2020fedhome}  & & & & &\\
        \hline
        Qayyum & \cmark & \xmark & \cmark & \xmark & \xmark\\
         et al. \cite{qayyum2022collaborative}  & & & & &\\
        \hline
       Liu & \cmark & \xmark & \xmark & \xmark & \xmark\\
         et al. \cite{liu2022contribution}  & & & & &\\
        \hline
       Sakib  & \cmark & \xmark & \xmark & \cmark & \xmark\\
         et al. \cite{sakib2021asynchronous}  & & & & &\\
        \hline
       Proposed & \cmark & \cmark & \cmark & \cmark & \cmark\\
      work  & & & & &\\
        \hline
    \end{tabular}
    \label{tab:1}
\end{table}

\section{GAN and FL-based Heart Health Monitoring}
\label{pro}
\begin{figure*}
    \centering
    \includegraphics[width=0.9\linewidth, height=3.2in]{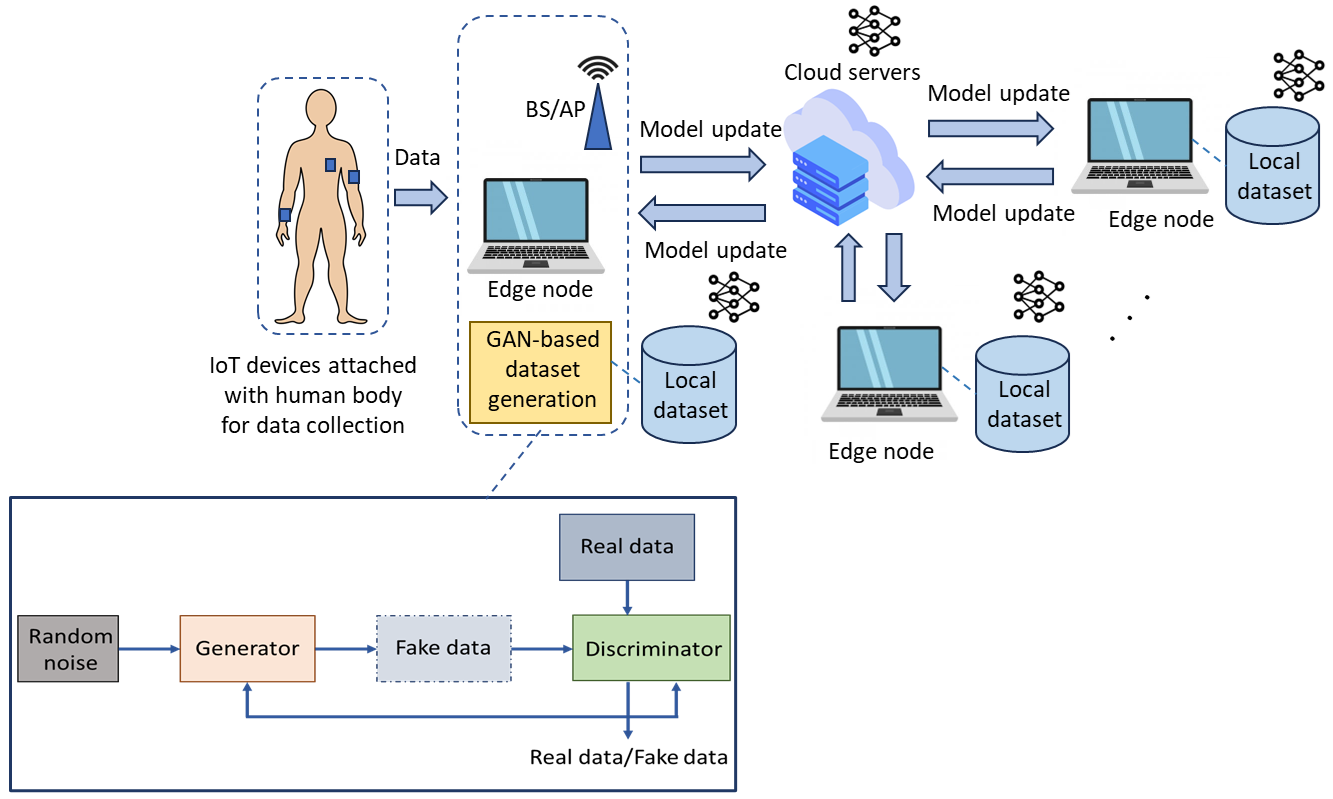}
    \caption{Proposed IoHT system model with GAN and CFL}
    \label{sys_cfl}
\end{figure*}

In this section, we first discuss the system model, then the use of GAN for generating realistic synthetic data, and finally, we discuss the use of CFL and DFL for developing the personalized and global health monitoring models. The proposed system model with GAN-based dataset generation and CFL is presented in Fig. \ref{sys_cfl}. The DFL architecture with three edge nodes is shown in Fig. \ref{dflpic}. The mathematical symbols are summarized in Table \ref{notation}.

\begin{table}
    \centering
    \caption{Mathematical symbols used in the proposed work}
    \begin{tabular}{c|c}
    \hline
       Notation & Meaning\\
       \hline
       $G$  &  Generator\\
       $D$ & Discriminator\\
       $G_{in}$ & Initial generator\\ 
       $D_{in}$ & Initial discriminator\\
       $l$ & Latent space\\
       $m$ & Number of data instances\\
       $n$ & Noise\\
       $\mathcal{D}_r$ & Real data\\
       $\mathcal{D}_f$ & Fake data\\
       $\mathcal{B}_1$ & Batch size in data generation process\\
       $\mathcal{L}_D$ & Discriminator's loss\\
       $\mathcal{L}_G$ & Loss in data generation\\
       $\eta_1$ & Number of epochs in data generation process\\
       $\gamma_g$ & Learning rate in data generation process\\
       $G_{f}$ & Final generator after training\\ 
       $D_{f}$ & Final discriminator after training\\
       $M_{in}$ & Initial Model\\
       $M_{s}$ & Model of the server\\
       $M_k$ & Model of edge node $k$\\
       $M_{final}$ & Final global model in CFL\\
       $M_{g}$ & Global model in DFL\\
       $\omega_{in}$ & Initial model weight\\
       $\omega_{s}$ & Model weight of the server\\
       $\omega_k$ & Model weight of edge node $k$\\
       $\omega_{final}$ & Final global model weight in CFL\\
       $\omega_{g}$ & Global model weight in DFL\\
       $\xi$ & Number of epochs in FL\\
       $\phi$ & Learning rate in FL\\
       $\mathcal{D}_k$ & Local dataset of edge node $k$\\
       $\mathcal{B}$ & Batch size in FL\\
       $K$ & Number of edge nodes\\
       $\alpha$ & Fraction of edge nodes participating in each round of CFL\\
       $N_k$ & Number of neighbours of edge node $k$ in DFL\\
       $R$ & Number of rounds in FL\\
       $T_{in}$ & Time consumption in model initialization\\
       $T_{loc}$ & Time consumption in local model training\\
       $T_{agg}$ & Time consumption in aggregation of model weights\\
       $T_{FL}$ & Time consumption in FL process\\
       $E_{CFL}$ & Energy consumption in CFL\\
       $E_{DFL}$ & Energy consumption in DFL\\
       $E_{edge}$ & Energy consumption of edge nodes\\
       $E_{cloud}$ & Energy consumption of cloud servers\\
       \hline
    \end{tabular}
    \label{notation}
\end{table}

\subsection{Proposed IoHT System}
The proposed IoHT system consists of (i) IoT devices used for collecting health data such as heart rate, blood pressure, and blood sugar, (ii) edge nodes where the collected data are stored and analyzed, and (iii) cloud servers. The edge node can be a mobile device, such as a smartphone, tablet, laptop, or any other electronic gadget with sufficient storage and processing ability to locally analyze the data. In a conventional IoHT system, the data are stored and analyzed inside the server. However, continuous data transmission may not be always feasible due to connectivity interruption as well as it increases the network traffic. Further, health data is confidential. Thus, the storage and analysis of health data inside the cloud may compromise with data privacy. To deal with these issues, we have used GAN, FL, and edge computing in our system. In the proposed system, we have considered edge nodes with high processing and storage ability so that they are capable enough to generate realistic synthetic data from the collected dataset, and then analyze the data locally through a collaborative learning process using FL. We assume that the edge node is connected to the cloud servers through an access point (AP) or base station (BS). 

\subsection{GAN-based Dataset Generation}
To address the challenges of data scarcity and class imbalance, the GAN is used to generate realistic synthetic data. The generated data resembles the original data. The GAN-based process for synthetic dataset generation is stated in Algorithm \ref{algo:gan_training}. The principal elements of a GAN are the generator and the discriminator. The generator is a deep neural network (DNN) where the random noise is fed as the input, and then it is converted to complex data samples that resemble the real data. The discriminator is an artificial neural network (ANN) that differentiates between real data and generated data samples. As the similarity between the generated data and real data samples increases, the generator's loss is reduced. As the accuracy of differentiating generated samples from the real samples increases, the discriminator's loss is reduced. The objective of a GAN is to maximize the discriminator's loss and minimize the generator's loss. Hence, the objective function of a GAN is mathematically presented as \cite{goodfellow2020generative}:
\begin{equation*}
\begin{split}
\min_G\max_D(G, D)=\mathbb{E}_{x \sim {p_{\mathcal{D}}(x)}}[log D(x)]\\ + \mathbb{E}_{n\sim p_{n}(n)}[log(1-D(g(n)))]
 \end{split}
\end{equation*}

where $G$ and $D$ represent the generator and discriminator networks respectively, $p_\mathcal{D}(x)$ denotes the probability distribution of original data, $p_n(n)$ denotes the probability distribution of the latent code or noise, $D(x)$ denotes the likelihood of $D$ to identify original data correctly, and $D(G(n))$ denotes the likelihood of $D$ to identify the authentic generated data. 

\begin{algorithm}
\caption{GAN-based algorithm for synthetic dataset generation}
\label{algo:gan_training}
\small
\begin{algorithmic}[1]
  \renewcommand{\algorithmicrequire}{\textbf{Input:}}
  \renewcommand{\algorithmicensure}{\textbf{Output:}}
    \Require $\mathcal{D}_r$, $\eta_1$, $\mathcal{B}_1$, $l$ 
  \Ensure $G_f$, $D_f$, $\mathcal{D}_f$

  \State $G_{in} \gets init\_Generator()$
  \State $D_{in} \gets init\_Discriminator()$
  
  \For{$i=1 \; to \; \eta_1$}
    \For{$b \in \mathcal{D}_b \subset \mathcal{D}_{r}$}
      \State Sample noise $n^{(1)}, n^{(2)}, \dots, n^{(m)}$ from $l$
      \State $\mathcal{D}_f \gets G(n; G_{in})$ 
      \State $D_r \gets \Delta(\mathcal{D}_b; D_{in})$ \Comment{discriminator output for real data}
      \State $D_f \gets \Delta(\mathcal{D}_f; D_{in})$ \Comment{discriminator output for fake data}
      \State 
      $\mathcal{L}_{D} \gets -\frac{1}{b} \sum_{j=1}^{b} \left[ \log D_r^{(j)} + \log (1 - D_f^{(j)}) \right]$ \Comment{compute discriminator loss}
      \State 
      $D_{in} \leftarrow D_{in} - \gamma_g \nabla_{D_{in}} \mathcal{L}_D$ \Comment{update discriminator parameters}
      \State Sample new noise $n^{(1)}, n^{(2)}, \dots, n^{(m)}$ from $l$
      \State  $D_f \gets \Delta(G(n; G_{in}); D_{in})$
      \State $\mathcal{L}_G \gets -\frac{1}{b} \sum_{j=1}^{b} \log D_f^{(j)}$ 
      \State $G_{in} \leftarrow G_{in} - \gamma_g \nabla G_{in} \mathcal{L}_G$ \Comment{update generator parameters}
    \EndFor
  \EndFor
  \State $G_f \gets G_{in}$
  \State $D_f \gets D_{in}$
  \State return $G_f, D_f$, $\mathcal{D}_f$
\end{algorithmic}
\end{algorithm}

\subsection{FL-based Heart Health Monitoring}
In the proposed work, we use CFL as well as DFL. The client and server-side processes of the CFL are stated in Algorithms \ref{algo_1} and \ref{algo_2} respectively. In Fig. \ref{sys_cfl}, the CFL process is presented, where edge nodes exchange model updates with the cloud server. In the CFL-based system, the edge nodes working as the clients receive model updates from the server, train their local models using their individual generated datasets, and send the local model updates to the server. For a number of rounds, the process is repeated till the loss becomes minimal. At the end of the process, each node has a personalized model as well as a global model. We have used GRU as the underlying DL model for analyzing the data, because GRU is faster, consumes less memory, and is suitable for sequential data.  
\begin{algorithm} 
\caption{Client-side algorithm}
\label{algo_1}
\begin{algorithmic}[1]
  \renewcommand{\algorithmicrequire}{\textbf{Input:}}
   \renewcommand{\algorithmicensure}{\textbf{Output:}}
  \Require $\mathcal{D}_k$, $\mathcal{B}$, $\xi$
  \Ensure $M_k$\\
  \textbf{$Update(\mathcal{D}_k, \mathcal{B}, \xi)$}:
  \State{$\mathcal{D}_k \gets preprocess(\mathcal{D}_k)$} 
  \While{($Server\_is\_connected == TRUE$)}
       \State{$M_k \gets receive()$}
       \State{$Train(\omega_k \gets M_k \cdot getweight())$}
    \EndWhile
    \State{$save(\omega_k)$}\\
   \textbf{$Function\;Train(\omega_k)$}:
      \State{$\beta \gets split(\mathcal{D}_k, \mathcal{B})$}    
      \For{$q=1 \; to \; \xi$} 
           \For{$b=1\; to \; \beta$}
              \State{$\omega_k^{q+1} \gets \omega_k^{q}-\phi\nabla \omega_k^q$}
           \EndFor
    \EndFor
    \State{$\omega_k \gets \omega_k^{\xi+1}$}
    \State{$M_k.setweight(\omega_k)$}
    \State{$send(M_k)$} 
\end{algorithmic}
\end{algorithm}

\begin{algorithm} 
\caption{Server-side Algorithm}
\label{algo_2}
\begin{algorithmic}[1]
  \renewcommand{\algorithmicrequire}{\textbf{Input:}}
   \renewcommand{\algorithmicensure}{\textbf{Output:}}
  \Require $K$, $\xi$, $R$
  \Ensure $M_{final}$\\
  \textbf{$Collect(K, R)$}:
  \State{$Clients \gets \emptyset$}
  \While{$(|Clients| \neq K)$}
       \State{$listen()$}
       \State{$accept\_connection()$}
    \EndWhile
   \State{$M_{in} \gets initializeModel()$} 
   \State{$send(M_{in})$}
   \State{$\omega_{in} \gets M_{in}.getweight()$} 
    \State{$M_{s} \gets M_{in}$} 
   \State{$\omega_{s}^0 \gets \omega_{in}$} 
   \State{$Aggregate()$}
    \State{$release\_clients(K)$} \\
      \textbf{$Aggregate()$}:
      \For{$r =0 \; to \; R-1$}
              \State{$C_r \gets Subset(max(\alpha*K,1),``random")$}
              \State{$W \; \gets \; []$}
             \For{$k \in C_r$}
             \State{$M_k \gets receive()$}
             \State{$\omega_k^r \gets M_k.getweight()$}
            \State{$W.append(\omega_k^r)$}
                   \EndFor
                   \State{$\omega_s^{r+1} \gets \frac{1}{|C_r|} \sum_{k=1}^{|C_r|} \cdot {\omega_k^r}$} \Comment{aggregation of received updates}
                   \State{$M_s.setweight(\omega_s^{r+1})$}
                   \State{$send(M_s)$}
    \EndFor
    \State{$\omega_{final} \gets \omega_s^{R}$}
    \State{$M_{final} \gets M_s$}
\end{algorithmic}
\end{algorithm}

\begin{figure}
    \centering
    \includegraphics[width=0.99\linewidth, height=1.8in]{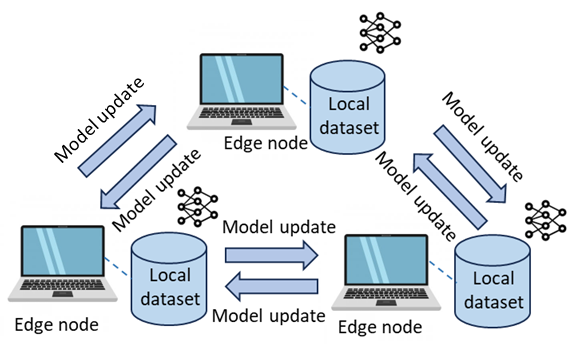}
    \caption{DFL framework with three edge nodes}
    \label{dflpic}
\end{figure}

For the DFL scenario, we consider the mesh topology in the proposed framework. The model update process in a mesh-based DFL network is stated in Algorithm \ref{algo_3}. A DFL network with three edge nodes is presented in Fig. \ref{dflpic}. In the DFL process, the edge nodes receive the initial model from the server, and then form a network among themselves to exchange model updates. Here, each node trains its local model using its locally generated dataset and sends the model update to the rest of the nodes. For a number of rounds, this process is repeated till a global model with minimal loss is obtained. At the end of the process, each node has a personalized model as well as a global model. 
\begin{algorithm}
\caption{Model update in Mesh-based DFL}
\label{algo_3}
\begin{algorithmic}[1]
\renewcommand{\algorithmicrequire}
{\textbf{Input:}}\renewcommand{\algorithmicensure}
{\textbf{Output:}}\Require $\mathcal{D}_k$, $N_k$, $R$, $\mathcal{B}$, $\xi$
\Ensure $M_k$, $M_g$ \\
\textbf{$Update(N)$}:
\For{$k =1 \; to \; K$}
\State $M_{in} \gets getinitialModelParamfromServer()$
\State{$\omega_{in} \gets M_{in}.getweight()$}
\State{$M_k \gets M_{in}$}
\State{$\omega_k^0 \gets \omega_{in}$}
\EndFor
\State{$M_g \gets M_{in}$}
\For{$r=0 \; to \; R-1$}
\For{$k = 1 \; to \; K$}   
\State{$\mathcal{D}_k \gets preprocess(\mathcal{D}_k)$}
\State{$\beta \gets split(\mathcal{D}_k, \mathcal{B})$}    
      \For{$q =1 \; to \; \xi$} 
           \For{$b=1 \; to \; \beta$}
           \State{$\omega_k^{q+1} \gets \omega_k^{q}-\phi\nabla \omega_k^q$}
           \EndFor
    \EndFor 
\State{$W \gets []$}
\For{$u = 1 \; to \; N_k$} 
    \State Send $\omega_k$ to neighbour $u$
    \State $\omega_u^r \gets receivefromneighbour(u)$
    \State{$W.append(\omega_u^r)$}
    \EndFor
    \State $\omega_k^{r+1} \gets \sum_{u=1}^{N_k} \omega_u^r/N_k$  \Comment{aggregation of received updates}
\EndFor
\State{$\omega_g^{r+1} \gets \sum_{k=1}^{K} \omega_k^{r+1}/K$}
\State{$M_g.setweight(\omega_g^{r+1})$}
\EndFor
\end{algorithmic}
\end{algorithm}

\subsection{Computational Complexity}
In this section, we discuss the time and computational complexity of the proposed approach. 
\begin{itemize}
    \item Time complexity of data generation process: The time complexity of the data generation process is $O(\eta_1 \cdot (\mathcal{D}/\mathcal{B}_1) \cdot (\omega_{G} + \omega_{D}))$, where $\omega_G$ and $\omega_D$ presents the model weights of the generator and discriminator respectively, $\mathcal{D}$ denotes data size, and $\mathcal{B}_1$ denotes batch size. The computational complexity of the process is also $O(\eta_1 \cdot (\mathcal{D}/\mathcal{B}_1) \cdot (\omega_{G} + \omega_{D}))$. 
    \item Time complexity of CFL process: The time complexity of initializing the model in CFL is $O(1)$. The computational complexity of model initialization is $O(\omega_{in})$, where $\omega_{in}$ is the initial model weight. The time complexity of receiving the initial model parameters from the server is $O(\omega_{in})$. The computational complexity of receiving the initial model parameters from the server is $O(K \cdot \omega_{in})$. 
    The time complexity of training a local model in CFL is $O(\omega_k \cdot \beta \cdot \xi \cdot R)$. The computational complexity of training a local model is $O(\omega_k \cdot \beta \cdot \xi \cdot R)$. 
    The time complexity of model updates' transmission in CFL is $O((\omega_k + \omega_s) \cdot R)$. The computational complexity of model updates' transmission is $O((K \cdot \omega_k + \omega_s )\cdot R)$. 
    The time complexity of aggregation in CFL is $O(\omega_k \cdot R \cdot K)$. The computational complexity of aggregation is $O(\omega_k \cdot R \cdot K)$. 
    \item Time complexity of DFL process: The time complexity of model initialization by the server is $O(1)$. The computational complexity of model initialization by the server is $O(\omega_{in})$, where $\omega_{in}$ is the initial model weight. The time complexity of receiving the initial model parameters from the server is $O(\omega_{in})$. The computational complexity of receiving the initial model parameters from the server is $O(K \cdot \omega_{in})$. 
    The time complexity of training a local model in DFL is $O(\omega_k \cdot \beta \cdot \xi \cdot R)$. The computational complexity of training a local model is $O(\omega_k \cdot \beta \cdot \xi \cdot R)$. 
    The time complexity of model updates' transmission in DFL is $O(\omega_k \cdot R)$. The computational complexity of model updates' transmission is $O(N_k \cdot \omega_k \cdot R)$. 
    The time complexity of aggregation by each node in DFL is $O(\omega_k \cdot R \cdot N_k)$. The computational complexity of aggregation by each node in DFL is $O(\omega_k \cdot R \cdot N_k)$. The time complexity of overall aggregation in DFL is $O(\omega_k \cdot R \cdot K)$. The computational complexity of overall aggregation in DFL is $O(\omega_k \cdot R \cdot K)$.
\end{itemize}

\subsection{Total Time in FL}
Total time in FL is given as,
\begin{equation}
\label{flt}
    T_{FL} = T_{in} + T_{loc} + T_{exc} + T_{agg}
\end{equation}
where $T_{in}$, $T_{loc}$, $T_{exc}$, and $T_{agg}$ denote time consumption in model initialization, local model training, exchange of model updates, and aggregation, respectively.

\subsection{Energy Consumption in FL}
The energy consumption in CFL is determined by summing up the energy consumption of the edge nodes ($E_{edge}$) and the cloud ($E_{cloud}$), as follows:
\begin{equation}
\label{fl}
    E_{CFL} = E_{edge} + E_{cloud}
\end{equation}
where
\begin{equation}
\label{es}
     E_{edge} = \sum_{k=1}^{K} \sum_{r=1}^{R}(T_{loc_k}+T_{exc_k}) \cdot e_{edge}
\end{equation}
\begin{equation}
\label{ec}
    E_{cloud} = (T_{in} \cdot e_{cloud}) + \sum_{r=1}^{R}(T_{excs} + T_{agg}) \cdot e_{cloud}
\end{equation}
where $e_{edge}$ and $e_{cloud}$ denote the power consumption of an edge node and the cloud per unit time, respectively, $T_{in}$ is the time consumption in model initialization by the cloud, and $T_{loc_k}$, $T_{exc_k}$, $T_{excs}$, and $T_{agg}$ denote the time consumptions in local model training for edge node $k$, model updates exchange for node $k$, model updates exchange for the cloud, and aggregation of model updates per round, respectively. 
\par
The energy consumption in DFL is determined as follows:
\begin{equation}
\label{es}
     E_{DFL} = (T_{in} \cdot e_{cloud}) + \sum_{k=1}^{K} \sum_{r=1}^{R}(T_{loc_k}+T_{exc_k}+T_{agg}) \cdot e_{edge} 
\end{equation}
where $e_{edge}$ and $e_{cloud}$ denote the power consumption of an edge node and the cloud per unit time, respectively, $T_{in}$ is the time consumption in model initialization by the cloud, and $T_{loc_k}$, $T_{exc_k}$, and $T_{agg}$ denote the time consumptions in local model training, model updates exchange, and aggregation of model updates for node $k$ per round, respectively. 

\subsection{Response Time}
The response time is calculated as the difference between the time stamps of submitting a request for prediction and receiving the result. When a request is received, the respective edge node performs prediction using the model and generates the result. The response time calculated as,
\begin{equation}
\label{response}
    T_{resp} = T_{res} - T_{req} 
\end{equation}
where the time stamp of submitting the request is $T_{req}$ and the time stamp of receiving the result is $T_{res}$.

\section{Performance Evaluation}
\label{per}
To implement the proposed work, we used Python 3.8.10. Tensorflow is used for implementing the DL models. For implementing CFL, a client-server model is developed, and MLSocket is used for communication between the server and clients. For implementing DFL, four nodes are connected using mesh topology, and the communication among the nodes is established using MLSocket. The number of rounds in FL is considered five and the number of epochs per round is considered fifty.

\subsection{Experimental Setup and GAN-based Dataset}
For the experimental setup, we have procured five virtual machines (VMs) from the Amazon AWS RONIN Cloud environment. Among these five VMs, four VMs are used as the client machines, and one is used as the server machine. Each machine has 4GB memory, 2vCPUs, and 100 GB SSD. From the original heart health monitoring dataset\footnote{\url{https://www.kaggle.com/datasets/johnsmith88/heart-disease-dataset}}, 1000 samples are used to generate the synthetic dataset using GAN. The generated synthetic dataset contains 10,000 samples, which are non-identically split among the four client machines. Randomly selected 400 samples from the original dataset are assigned as the global dataset to the server. Randomly selected 500 samples from the original dataset are used as the test dataset for each client. 
\par
The statistical summary of the generated dataset are presented in Table \ref{gandata}. The health parameters used as the input parameters are presented in the table. The output is the target class (0 (no disease) or 1 (disease)). The maximum value, minimum value, mean, and standard deviation (SD) of the parameters of the generated datasets are presented in the table.

\begin{table*}
    \centering
    \caption{\textcolor{black}{Statistical summary of the generated dataset}}
    \begin{tabular}{ c c c c c }
    \hline
        Parameter & Minimum value & Maximum value & Mean & SD\\ 
       
    \hline
       age & 32 & 76 & 55 & 6.14\\
       sex & 0 & 1 & 0.73 & 0.44\\
       resting blood pressure (trestbps) & 98 & 200 & 126 & 12.63\\
       chest pain type (cp) & 0 & 3 & 0.48 & 0.79\\
       serum cholesterol (chol) & 126 & 409 & 256 & 32.67\\
       oldpeak & 0 & 4.4 & 0.54 & 0.56\\
       resting electrocardiographic results (restecg) & 0 & 2 & 0.02 & 0.15\\
       fasting blood sugar $>$ 120 mg/dl (fbs) & 0 & 1 & 0.01 & 0.09\\
       exercise-induced angina (exang) & 0 & 1 & 0.22 & 0.41\\
       maximum heart rate achieved (thalach) & 81 & 199 & 158 & 21.04\\
       number of major vessels (0-3) colored by fluoroscopy (ca) & 0 & 4 & 0.5 & 0.82\\
       thal & 0 & 3 & 2.41 & 0.53\\
       the slope of the peak exercise ST segment (slope) & 0 & 2 & 1.44 & 0.5\\
    \hline
    \end{tabular}
    \label{gandata}
\end{table*}

\begin{figure*}
    \centering
    \includegraphics[width=0.99\linewidth, height=5.8in]{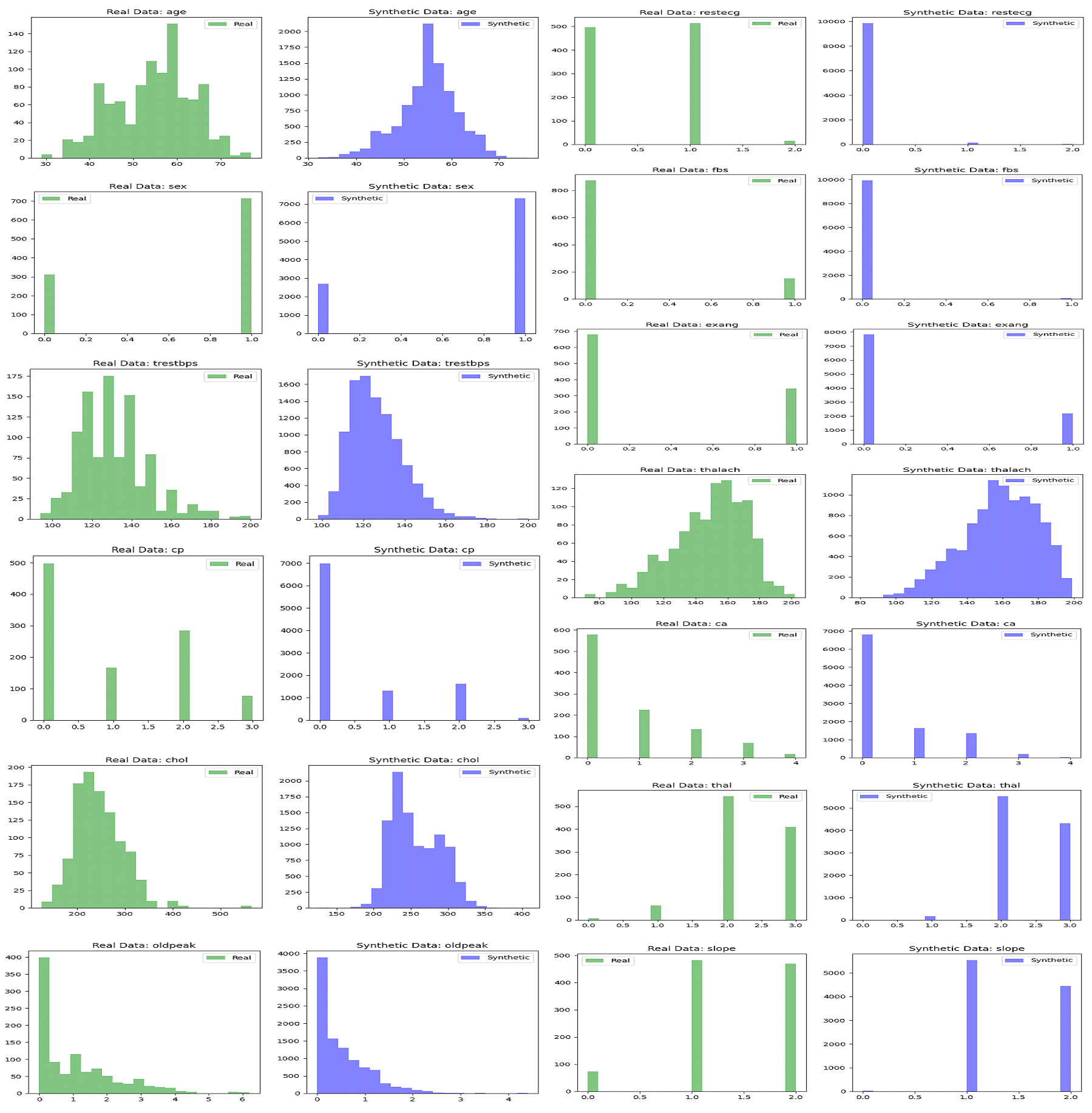}
    \caption{\textcolor{black}{Histogram analysis of the real and synthetic datasets}}
    \label{histoheart}
\end{figure*}

\textcolor{black}{For validation of the GAN-based dataset, we have performed histogram analysis, and presented the results in Fig. \ref{histoheart}. The histogram analysis of real and synthetic data is presented for the attributes, and we observe that the data distribution for the synthetic data matches that of the real dataset for most of the attributes.} 

\begin{figure}
    \centering
    \includegraphics[width=0.99\linewidth, height=0.8in]{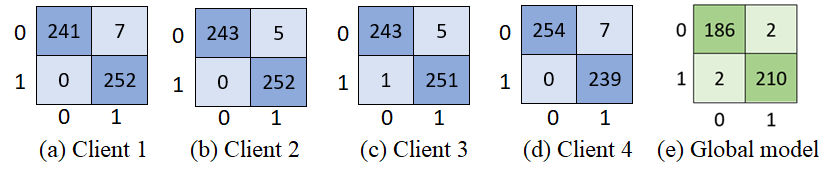}
    \caption{Confusion matrices in CFL for the test datasets}
    \label{con1}
\end{figure}

\subsection{CFL-based Model}
\begin{figure}
    \centering
    \includegraphics[width=0.99\linewidth, height=2.0in]{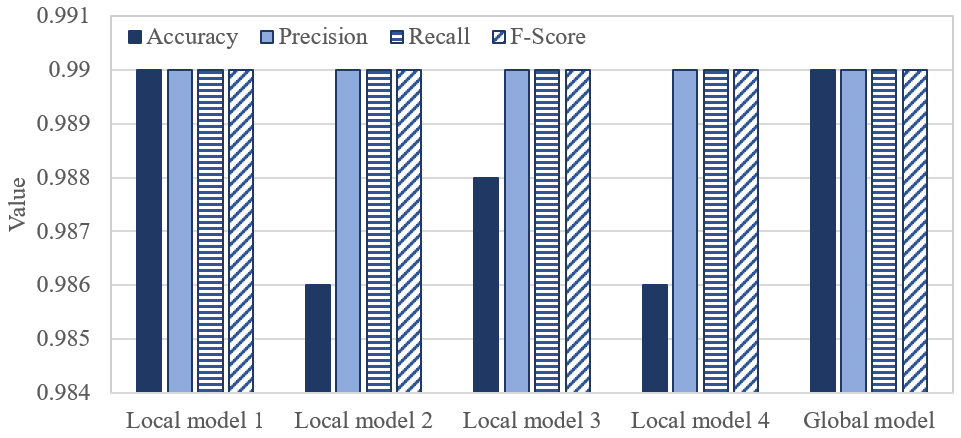}
    \caption{Accuracy metrics in CFL}
    \label{acc}
\end{figure}

\begin{figure}
    \centering
    \includegraphics[width=0.9\linewidth, height=2.0in]{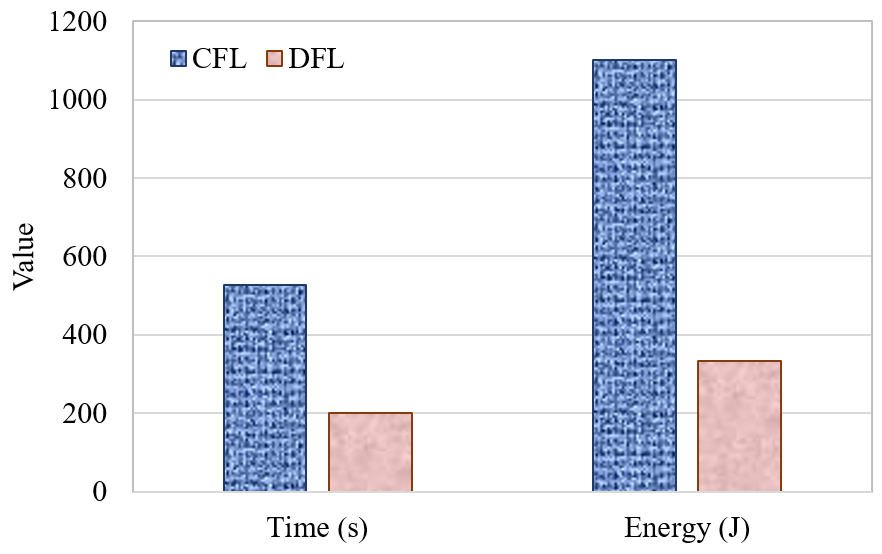}
    \caption{Time and energy consumption in CFL and DFL}
    \label{timenfl}
\end{figure}

\begin{figure}
    \centering
    \includegraphics[width=0.99\linewidth, height=2.5in]{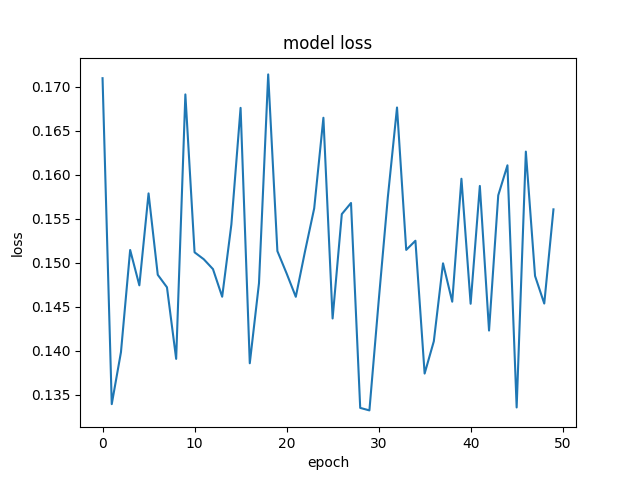}
    \caption{Loss of the global model in CFL}
    \label{loss_cfl}
\end{figure}

In CFL, the model updates are exchanged between the clients and the server to build the final global model. The confusion matrices for the local models and the global model are presented in Fig. \ref{con1}. The prediction accuracy, precision, recall, and F-score of the local and global models are presented in Fig. \ref{acc}. 
The global model achieves an accuracy of 0.99, and the precision, recall, and F-score are also 0.99. The average precision, recall, and F-score of the personalized local models are also 0.99. The average accuracy of the personalized local models is approximately 0.99. The total time and energy consumption in CFL are presented in Fig. \ref{timenfl}. The total time and energy consumption by the clients and the server considering all five rounds of FL, are 527s and 1100J, respectively. The global model loss after five rounds is presented in Fig. \ref{loss_cfl}. The loss is minimal at the convergence point. We observe that the loss is $<$0.17. \par

\subsection{DFL-based Model}
\begin{figure}
    \centering
    \includegraphics[width=0.99\linewidth, height=0.8in]{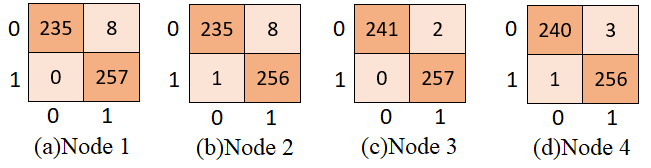}
    \caption{Confusion matrices in DFL for the test datasets}
    \label{conDFL}
\end{figure}

\begin{figure}
    \centering
    \includegraphics[width=0.99\linewidth, height=2.0in]{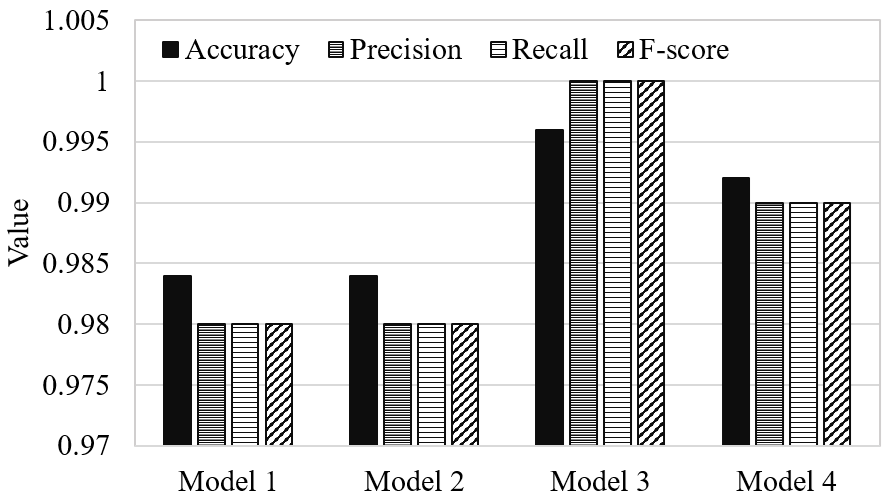}
    \caption{Accuracy metrics in DFL}
    \label{acc_dfl}
\end{figure}

\begin{figure}
    \centering
    \includegraphics[width=0.99\linewidth, height=2.5in]{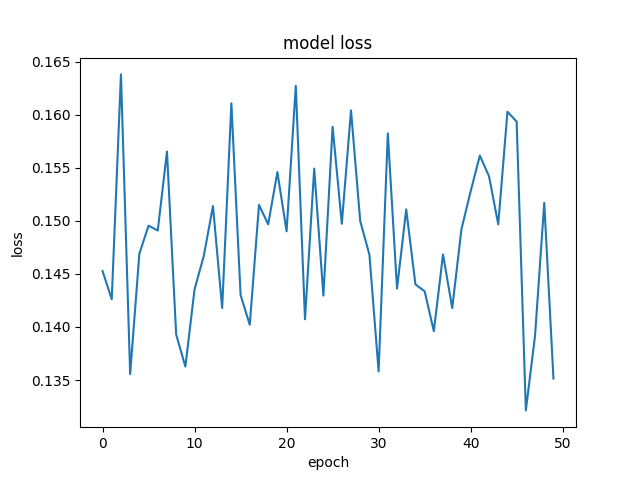}
    \caption{Loss of the global model in DFL}
    \label{loss_dfl}
\end{figure}

In the experiment, the local models of each node are trained using the GAN-generated dataset. Each node exchanges model updates with its neighbours to build the global model. The confusion matrices of the models of the participating nodes are presented in Fig. \ref{conDFL}. The accuracy, precision, recall, and F-score while using GRU-based DFL, are presented in Fig. \ref{acc_dfl}. For the DFL, the models of the participating nodes have achieved an average accuracy, precision, recall, and F-score of approximately 0.99. The total time consumption in DFL is 200s, considering all five rounds, as presented in Fig. \ref{timenfl}. The sum of the energy consumption of a VM for model initialization and the energy consumption of all four edge nodes, considering all five rounds is 334J, as presented in Fig. \ref{timenfl}. To evaluate the performance of the final global model in DFL, we use the global dataset. The prediction accuracy, precision, recall, and F-score of the global model in DFL are 0.99. The loss of the global model after round 5 while using DFL is presented in Fig. \ref{loss_dfl}. The loss is minimal at the convergence point. We observe that the loss is $<$0.165. As we observe from the results for both CFL and DFL, the proposed framework has achieved an accuracy, precision, recall, and F-score of $\sim99\%$.

\subsection{Comparison with the Conventional Framework}
\begin{figure}
    \centering
    \includegraphics[width=0.99\linewidth, height=2.5in]{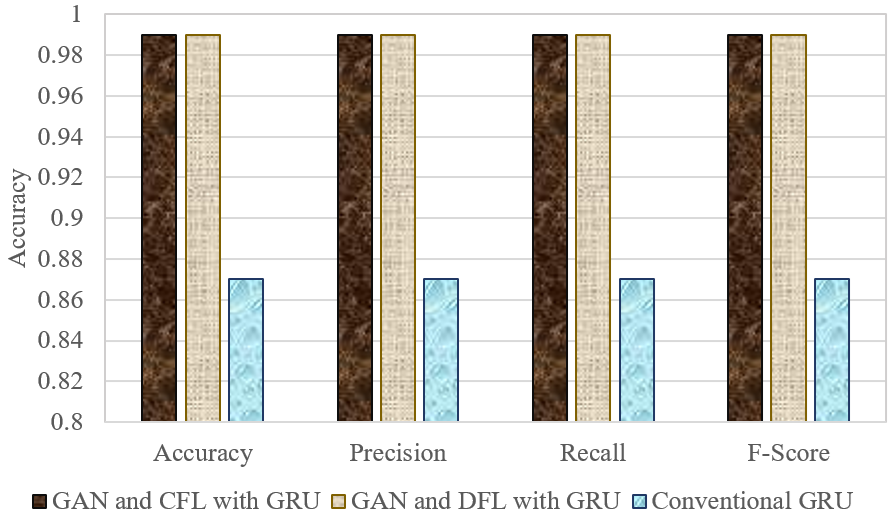}
    \caption{Comparison of prediction accuracy between the proposed work and conventional GRU-based framework}
    \label{comp}
\end{figure}

\begin{figure}
    \centering
    \includegraphics[width=0.99\linewidth, height=2.0in]{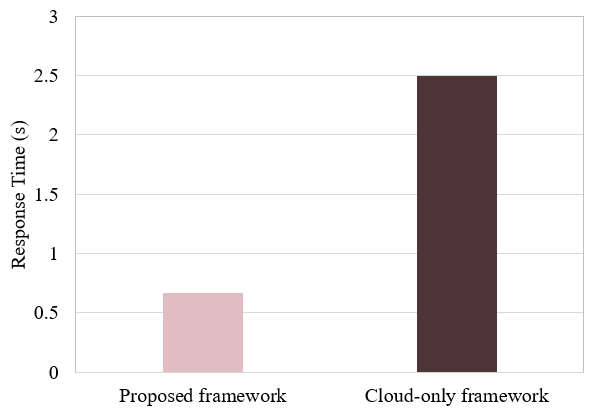}
    \caption{Comparison of response time between the proposed and cloud-only IoHT frameworks}
    \label{resp}
\end{figure}

We have compared the results with the conventional GRU-based framework. In this case, we have used the real dataset based on which the synthetic dataset was generated, and used GRU for training without FL. The accuracy, precision, recall, and F-score for this case are 0.89, which is much less than the proposed framework. The results are presented in Fig. \ref{comp}. As we observe the proposed GAN and FL-based framework has $\sim$12\% better accuracy than the conventional GRU-based framework without FL. We have also implemented the cloud-only IoHT framework and measured the response time for the cloud-only framework and the proposed FL-based framework. The average response time for the proposed FL-based framework is 0.67s, and the average response time for the cloud-only framework is 2.5s. The results are presented in Fig. \ref{resp}, and it is observed that the proposed framework has $\sim$73\% lower response time. 

\subsection{Comparison with Other ML and DL models}
\textcolor{black}{The accuracy metrics for the proposed framework are compared with other well-known ML and DL models. The results are presented in Table \ref{comp_othmodels}. From the results, we observe that the proposed model has higher prediction accuracy compared to the other ML and DL models, including Long Short-Term Memory (LSTM) network, Multilayer Perceptron (MLP), Conventional Neural Network (CNN), Support Vector Machine (SVM), and Random Forest (RF).} 
\begin{table} 
\caption{\textcolor{black}{Comparison of the proposed work with existing ML and DL models}}
    \centering
    \begin{tabular}{|c|c|c|c|c|}
        \hline
         Model   &  Accuracy &  Precision & Recall & F1-Score\\
          \hline
         LSTM & 0.84 & 0.84 & 0.84 & 0.84\\
         \hline
         MLP & 0.92 & 0.92 & 0.92 & 0.92\\
         \hline
         CNN & 0.96 & 0.96 & 0.96 & 0.96\\
         \hline
         SVM & 0.81 & 0.83 & 0.81 & 0.81\\
         \hline
         RF & 0.97 & 0.97 & 0.97 &0.97\\
         \hline
         Proposed & 0.99 & 0.99 & 0.99 & 0.99\\
         (FL with GAN) & & & &\\
        \hline
    \end{tabular}
    \label{comp_othmodels}
\end{table}

\subsection{Comparison with Existing Approaches} 
The performance of the proposed approach is compared with the existing schemes on FL-based heart healthcare, and heart disease detection.

\subsubsection{Comparison with FL-based heart healthcare} The proposed approach is compared with the existing FL-based heart healthcare approaches in Table \ref{tab:comp}. As we observe, the proposed approach has achieved a good prediction accuracy level. Further, in the proposed approach, we have implemented not only CFL but also DFL. For a hard-deadline application like healthcare, response time is crucial. For any learning process, time and energy consumption are also significant. Thus, we have not only measured the prediction accuracy level, but also time and energy consumption, while evaluating performance of the proposed framework. 

\begin{table} 
\caption{Comparison of the proposed work with the existing works on FL-based heart healthcare systems}
    \centering
    \begin{tabular}{|c|c|c|c|c|c|}
        \hline
         Work   &  Used & CFL/& Accuracy &  Total  &  Energy \\
         & classifier & DFL& & Time & \\
          \hline
         Yaqoob  & MBAC- &CFL & 93.8\%& Not & Not\\
         et al. \cite{yaqoob2023hybrid}  & SVM&& & measured& measured\\
         \hline
       Sakib   & CNN & CFL& 88-90\%& Not & Not\\
         et al. \cite{sakib2021asynchronous}   & & & & measured& measured\\
        \hline
       Proposed  & GRU & CFL,& $\sim$99\%& CFL: & CFL:\\
      work  & & DFL&&527s & 1100J\\
      & & &&DFL: & DFL: \\
      & & &&200s & 334J\\
        \hline
    \end{tabular}
    \label{tab:comp}
\end{table}

\subsubsection{Comparison with existing approaches on heart disease detection} Finally, we compare the proposed approach with the existing approaches on heart disease detection in Table \ref{comp2}. 
\begin{table*} 
\caption{Comparison of the proposed work with the existing works on heart disease detection}
    \centering
    \begin{tabular}{|c|c|c|c|c|c|c|c|}
        \hline
         Work   &  Used classifier & Used FL & Used GAN & Accuracy &  Precision  &  Recall & F-Score\\
          \hline
        Khan et al.\cite{khan2024utilising} & LR, SVM, KNN, DT, RF (Highest), GB & No & No & 85.25\% & 81.25\% & 89.66\% & 85.25\%\\
        \hline
        Praveen et al.\cite{praveen2024enhanced} & Integration of IQR outlier detection,
& No & No & 83\% & 84\% & 85\% & 84\%\\
& MICE, SMOTE, and GOL2-2T  & & & & & & \\
\hline
Abubaker et al.\cite{abubaker2023ensemble} & SVM, KNN (better), NB, EX, & No & No & 81\% & 81\% & 81\% & 81\%\\
& LDA, MLP, LR& & & & & & \\
\hline
Mir et al.\cite{mir2024novel} & KNN, DT, NB, RF, AdaBoost ensemble & No & No & 93.9\% & 92.23\% & 96.75\% & 94.44\%\\
& using RF (highest) & & & & & & \\
\hline
Bertsimas et al.\cite{bertsimas2021machine} & XGBoost & No & No & 93-99\% & 93-99\%  & 93-99\%  & 93-99\% \\
\hline
Proposed work & GRU & Yes & Yes & 99\% & 99\% & 99\% & 99\%\\
        \hline
    \end{tabular}
    \label{comp2}
\end{table*}
In \cite{khan2024utilising}, Logistic Regression (LR), SVM, K-Nearest Neighbour (KNN), Decision Tree (DT), RF, and Gradient Boost (GB), were used, and RF achieved the highest accuracy of 85.25\%. In \cite{praveen2024enhanced}, Synthetic Minority Over-sampling Technique (SMOTE), Inter Quartile Range (IQR) outlier detection, Multivariate Imputation by Chained Equations (MICE), and GOL2-2T, were used, and 83\% accuracy was achieved. In \cite{abubaker2023ensemble}, SVM, LR, KNN, Naïve Bayes (NB), MLP, Linear Discriminant Analysis (LDA), and Extra Tree (EX), were used, where KNN performed better, and achieved an accuracy of 81\%. In \cite{mir2024novel}, KNN, DT, NB, RF, and AdaBoost ensemble using RF were used, and AdaBoost ensemble using RF achieved the highest accuracy of 93.9\%. In \cite{bertsimas2021machine}, Extreme Gradient Boosting (XGBoost) was used for heart disease prediction, and the authors achieved an accuracy of 93\%-99\%. As we observe the existing approaches neither used GAN nor FL. We also observe that the proposed approach has accuracy, precision, recall, and F-score of $\sim99\%$, which is high than the existing schemes. Using GAN, a realistic dataset has been generated, and using FL the global model is developed. As we observe from the table, the proposed approach has outperformed the existing schemes.

\subsection{Clinical Relevance of the Study}
Heart health monitoring on a regular basis is vital to detect, treat, and prevent heart disorders, and improve the overall cardiovascular health. Continuous monitoring and detection of heart disease help to prevent serious complications. In the existing heart health monitoring systems, the heart-related data are stored and analyzed inside the cloud servers, which compromises with data privacy. On the other hand, accurate heart disease detection requires sufficient amount of data for proper training. However, the existing systems sometimes are not able to make accurate predictions due to data scarcity and class imbalance issues. To deal with the data privacy and data scarcity issues of the existing systems, the proposed work integrates GAN with FL, and proposes an accurate and secure heart health monitoring framework. The results also validate the proposed framework provides accurate heart disease detection at low response time, which makes it clinically relevant for the smart healthcare.

\section{Conclusions and Future Work}
\label{con}
This paper proposed a GAN and FL-based health monitoring system. Using the GAN, realistic synthetic datasets are generated to improve data diversity. The edge nodes have locally generated datasets, which are used to train the local models, and model updates are exchanged with the cloud servers for collaborative training. We also implement the DFL approach, where the edge nodes communicate among themselves to exchange model updates for collaborative training. As the underlying DL approach, GRU is used. From the results we observe that the proposed approach outperforms the existing approaches. The results present that using the proposed approach $\sim99$\% prediction accuracy is achieved, which is $\sim$12\% better than the conventional GRU-based framework. The results also demonstrate that the proposed framework lowers the response time by $\sim$73\% than the conventional cloud-only IoHT system. We also observe from the results that the proposed work outperforms the existing healthcare approaches. 
\par
The GAN-based dataset generation process require good processing ability of the device, and the mobile devices may not have sufficient processing capacity for its execution. Further, the large scale data analysis inside the mobile devices may slow down the process, which is known as the Straggler effect. In the future, we would like to explore lightweight dataset generation process and federated offloading to address these issues. We would also like to explore the use of bayesian deep learning approach for FL-based healthcare systems.

\bibliography{ref}
\bibliographystyle{IEEEtran}

\end{document}